\documentclass[driverfallback=dvipdfmx,usenatbib]{akari2017}
\usepackage{natbib}

\shorttitle{AKARI IR LFs}
\shortauthors{T. Goto}

\begin{document}

\title{Cosmic star formation history revealed by AKARI and Hyper Suprime-Cam } 




\author{Tomotsugu Goto} 
\affiliation{National Tsing hua University, No. 101, Section 2, Kuang-Fu Road, Hsinchu, Taiwan 30013} 
\author{Nagisa Oi} \affiliation{Tokyo University of Science, 1-3 Kagurazaka, Shinjuku-ku, Tokyo 162-8601, Japan}
\author{Rieko Momose}\affiliation{National Tsing hua University, No. 101, Section 2, Kuang-Fu Road, Hsinchu, Taiwan 30013} 
\author{Ece Kilerci Eser}\affiliation{National Tsing hua University, No. 101, Section 2, Kuang-Fu Road, Hsinchu, Taiwan 30013} 
\author{Hideo Matsuhara}\affiliation{Institute of Space and Astronautical Science, Japan Aerospace Exploration Agency, 3-1-1 Yoshinodai, Chuo, Sagamihara, Kanagawa 252-5210, Japan}
\author{Ting-Chi Huang}\affiliation{National Tsing hua University, No. 101, Section 2, Kuang-Fu Road, Hsinchu, Taiwan 30013} 
\author{Yousuke Utsumi}\affiliation{Kavli Institute for Particle Astrophysics and Cosmology (KIPAC), SLAC National Accelerator Laboratory, Stanford University, SLAC, 2575 Sand Hill Road, Menlo Park, CA  94025, USA}
\author{Yoshiki Toba}\affiliation{Academia Sinica Institute of Astronomy and Astrophysics, P.O. Box 23-141, Taipei 10617, Taiwan}
\author{Youichi Ohyama}\affiliation{Academia Sinica Institute of Astronomy and Astrophysics, P.O. Box 23-141, Taipei 10617, Taiwan}
\author{Toshinobu Takagi}\affiliation{Japan Space Forum, 3-2-1, Kandasurugadai, Chiyoda-ku, Tokyo 101-0062 Japan}
\author{Takehiko Wada}\affiliation{Institute of Space and Astronautical Science, Japan Aerospace Exploration Agency, 3-1-1 Yoshinodai, Chuo, Sagamihara, Kanagawa 252-5210, Japan}
\author{Matthew Malkan}\affiliation{Department of Physics and Astronomy, UCLA, Los Angeles, CA, 90095-1547, USA}
\author{Seong Jin Kim}\affiliation{National Tsing hua University, No. 101, Section 2, Kuang-Fu Road, Hsinchu, Taiwan 30013} 
\author{Takao Nakagawa}\affiliation{Institute of Space and Astronautical Science, Japan Aerospace Exploration Agency, 3-1-1 Yoshinodai, Chuo, Sagamihara, Kanagawa 252-5210, Japan}
\author{the AKARI team} 



 \begin{abstract}

 Understanding infrared (IR) luminosity is fundamental to
understanding the cosmic star formation history and AGN evolution,
since their most intense stages are often obscured by dust. Japanese
infrared satellite, AKARI, provided unique data sets to probe this
both at low and high redshift; the AKARI all sky survey in 6 bands
(9-160 $\mu$m), and the AKARI NEP survey in 9 bands (2-24$\mu$m).

  The AKARI performed all sky survey in 6 IR bands (9, 18, 65, 90,
140, and 160 $\mu$m) with 3-10 times better sensitivity than IRAS, covering
the crucial far-IR wavelengths across the peak of the dust emission.
Combined with a better spatial resolution, we measure the total infrared luminosity ($L_{TIR}$) of individual galaxies, and thus, the total infrared luminosity density of the local
Universe much more precisely than previous work.

 In the AKARI NEP wide field, AKARI has obtained deep images in the
mid-infrared (IR), covering 5.4 deg$^2$. However,
our previous work was limited to the central area of 0.25 deg$^2$ due
to the lack of deep optical coverage.
  To rectify the situation, we used the newly advent Subaru telescope's Hyper Suprime-Cam to obtain deep optical images over the entire 5.4 deg$^2$ of the AKARI NEP wide field.
  With this deep and wide optical data, we, for the first time, can use the entire AKARI NEP wide data to construct restframe 8$\mu$m, 12$\mu$m, and total infrared (TIR) luminosity
functions (LFs) at 0.15$<z<$2.2. A continuous 9-band filter coverage in
the mid-IR wavelength (2.4, 3.2, 4.1, 7, 9, 11, 15, 18, and 24$\mu$m) by
the AKARI satellite allowed us to estimate restframe 8$\mu$m and 12$\mu$m
luminosities without using a large extrapolation based on a SED fit,
which was the largest uncertainty in previous work.

  By combining these two results, we reveal dust-hidden cosmic star
formation history and AGN evolution from z=0 to z=2.2, all probed by
the AKARI satellite.

%
%
 \end{abstract}



\setcounter{page}{1}




\section{Local IR LF from AKARI all sky survey}
\label{142349_30Nov17}

Local infrared (IR) luminosity functions (LFs) are necessary benchmarks for high-redshift IR galaxy evolution studies.
 Any accurate IR LF evolution studies require accordingly accurate local IR LFs.

 We construct infrared galaxy LFs at redshifts of $z \leq 0.3$ from \textit{AKARI} space telescope, which performed an all-sky survey in six IR 
 bands (9, 18, 65, 90, 140 and 160 $\mu$m) with 3-10 times better sensitivity than its precursor IRAS.
 Availability of 160 $\mu$m filter is critically important in accurately measuring total IR luminosity of galaxies, covering across the peak of the dust emission. 
  By combining mid-IR data from Wide-field Infrared Survey Explorer (\textit{WISE}), and spectroscpic redshifts from
 Sloan Digital Sky Survey (SDSS) Data Release 13 (DR13), 6-degree Field Galaxy Survey (6dFGS) and the 2MASS Redshift Survey (2MRS), 
 we created a sample of 15,638 local IR galaxies with spectroscopic redshifts, by a factor of 20 larger compared with well-cited previous work based on IRAS data \citep{sanders2003}, which was also limited to $<$100 $\mu$m.

\begin{figure}
  \begin{center}
 \includegraphics[scale=0.75]{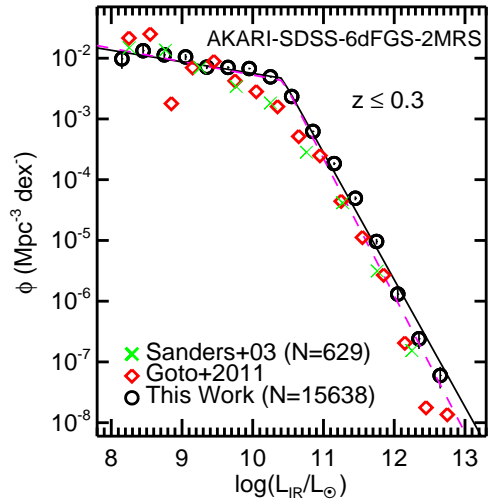}
    \caption{The IR LF of 15,638  \textit{AKARI$-$SDSS$-$6dFGS$-$2MRS}  galaxies (open circles). The best-fitting double power law is shown as solid line.
For comparison the total IR LF derived from the \textit{IRAS} RBGS is shown \citep[crosses][]{2003AJ....126.1607S}.The red diamonds are the $1/V_{max}$ data points of the RBGS sample adopted from \citet{2011MNRAS.410..573G}.The dashed magenta line is the best-fitting double power law when the RBGS data are included in the fit. }
  \label{fig:local_LF}
  \end{center}
 \end{figure}

 After carefully correcting for volume effects in both IR and optical,
 we show obtained IR LFs in Fig. \ref{fig:local_LF}, which agree well with previous studies, but comes with much smaller errors.  Especially both faint- and bright-ends of the LFs are better-determined, due to much larger size of the spectroscpic redshifts and the IR photometry.
 
 Measured local IR luminosity density is $\Omega_{IR}= 1.19\pm0.05 \times 10^{8} L_{\sun}$ Mpc$^{-3}$. 
 The contributions from luminous infrared galaxies and ultra luminous infrared galaxies to $\Omega_{IR}$ are very small, 9.3 per cent and 0.9 per cent, respectively.
 There exists no future all sky survey in far-infrared wavelengths in the foreseeable future.
 The IR LFs obtained in this work will therefore remain an important benchmark for high-redshift studies for decades. See more details in \citet{Ece2017}.

\section{High-z IR LFs from the AKARI NEP wide field}

\subsection{Undetected AKARI sources\label{sec:intro}}

The extragalactic background suggests at least half the energy generated by stars has been 
reprocessed into the infrared (IR) by dust \citep{1999A&A...344..322L}.
At z$\sim$1.3, 90\% of star formation is obscured by dust \citep{2005ApJ...632..169L,GotoTakagi2010,GotoCFHT}.
Therefore, a full understanding of the cosmic star formation
history inevitably needs an IR perspective, especially at high redshifts.

The AKARI space telescope has performed a deep mid-infrared imaging survey in the NEP region \citep{2009PASJ...61..375L}.
We are studying the multi-band data of these mid-IR galaxies as shown in Table 1 \citep{Takagi_PAH,GotoTakagi2010}. However, because  very dusty objects cannot be detected in the relatively shallow CFHT imaging data \citep[$r<$25.9ABmag;][]{2014A&A...566A..60O}, there remain 11,000 AKARI sources undetected in the optical. 
As a result, we lack understanding of the redshift and IR luminosity of these sources, i.e., they have been excluded from the past cosmic star formation history (CSFH) analysis. These sources could change our view of CSFH---if they all lie at 1$<z<$2, they will $double$ the cosmic star formation density at that epoch. 

\subsection{Uniqueness of AKARI mid-IR data\label{sec:uniq}}
The AKARI NEP is one of the best fields for this investigation,
due to the availability of continuous 9-band mid-IR filters. Spitzer lacks filters between 8 and 24$\mu$m  (the critical wide gap between IRAC and MIPS, excluding the tiny IRS peak up array at 16 $\mu$m). Similarly WISE also has a wide gap between 4 and 12$\mu$m filters. Therefore, no other telescope can provide continuous 9-band photometry in mid-IR wavelength (2-24$\mu$m) over 5.4 deg$^2$, until JWST performs a similar survey. JWST will also require a large amount of telescope time to survey 5.4 deg$^2$.  
AKARI's continuous  9-band photometry works as a low-resolution spectrum, 
which is critically important for the following key aspects:

\begin{itemize} \item  
 Two physical processes produce the mid-IR emission: hot dust around an AGN, and PAH emission from star-formation. Quantitatively separating these is of fundamental importance. The continuous 9 filters of AKARI have made this possible through precise SED fitting \citep[See examples in ][]{Takagi_PAH, 2014ApJ...784..137K}. Importantly, this is independent of extinction.
 \item 

 Accurately measuring the mid-IR emission line strength (PAH 7.7 $\mu$m) and continuum luminosity. 
Using the 9-band photometry as a low-resolution spectra, 
Ohyama et al. (2017, submitted) demonstrated that photometric PAH 7.7 $\mu$m line measurements agree well with spectroscopic ones. 

\end{itemize}
 Neither of these is possible if there is a large gap between mid-IR filters. Therefore, AKARI NEP is the only field, where the two different astrophysical 
power sources can be separated for thousands of IR galaxies, including those with heavy extinction.

The AKARI NEP also has been thoroughly observed in every other available waveband (Table 1), making it one of the premier large deep fields on the sky.
Its large area overcomes the serious problem of cosmic variance, which hampered previous IR CSFH studies. 
In particular, Spitzer's CDFS field was only $~0.25$ deg$^2$  \citep{2005ApJ...632..169L}, and measured an IR luminosity density nearly a factor of 10 different from other Spitzer fields \cite[e.g.,][]{2006MNRAS.370.1159B}. For the same reason, the single Suprime-Cam pointing in the center of the NEP deep field ($0.25$ deg$^2$) is not wide enough.
A large volume coverage also allows us to study environmental effects on galaxy evolution  \citep{2008MNRAS.391.1758K,cluster_LF}.
AKARI was a survey telescope, which observed 5.4 deg$^2$ in NEP using $\sim10$\% of the entire pointed observations available throughout the lifetime of the mission,  providing uniquely precious space-based IR data spanning a large enough area to overcome cosmic variance.

\subsection{Subaru Hyper Suprime-Cam Observation\label{sec:hsc}}
 \begin{figure}
  \begin{center}
  \includegraphics[scale=0.4]{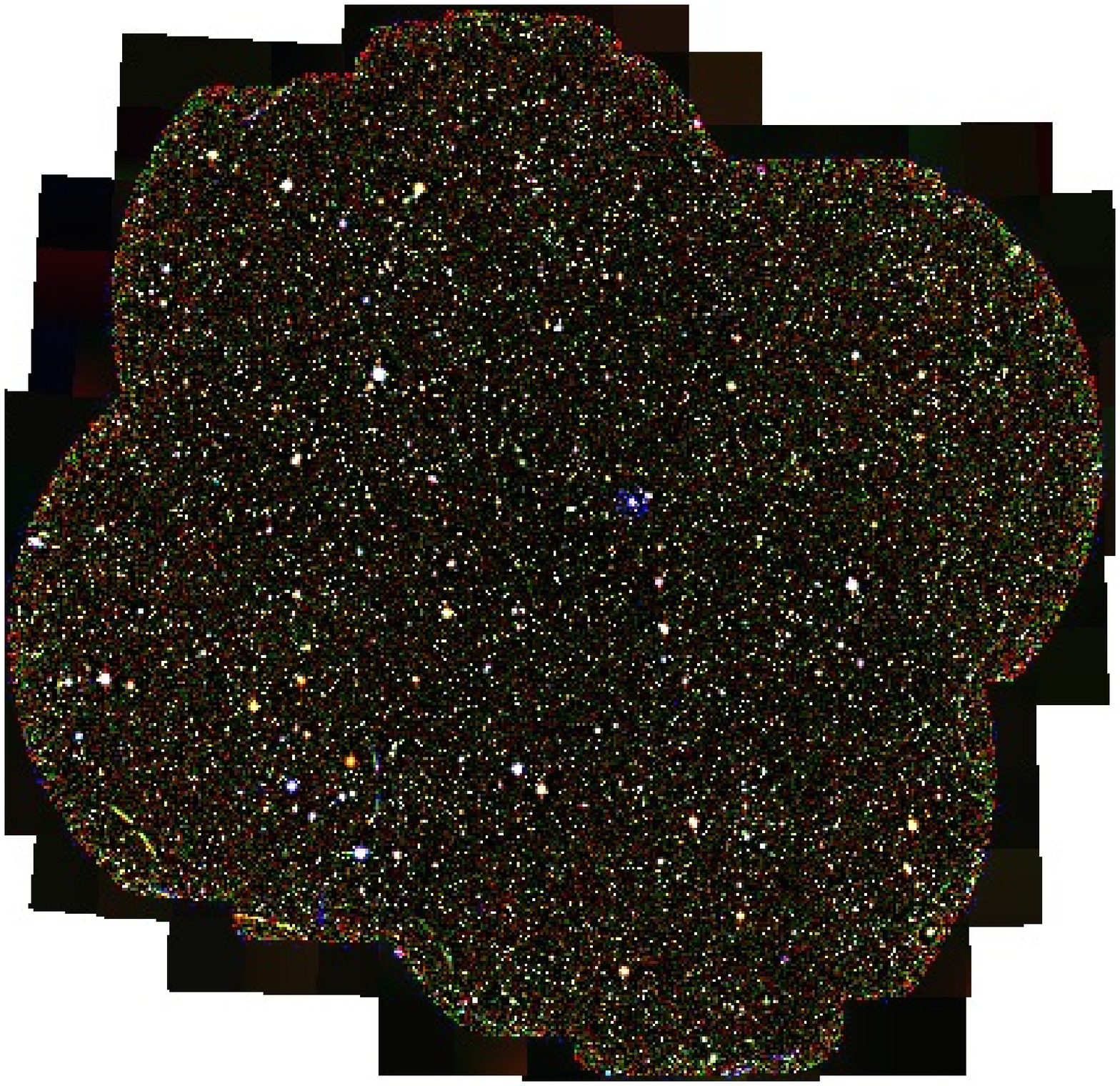}
  \includegraphics[scale=0.46]{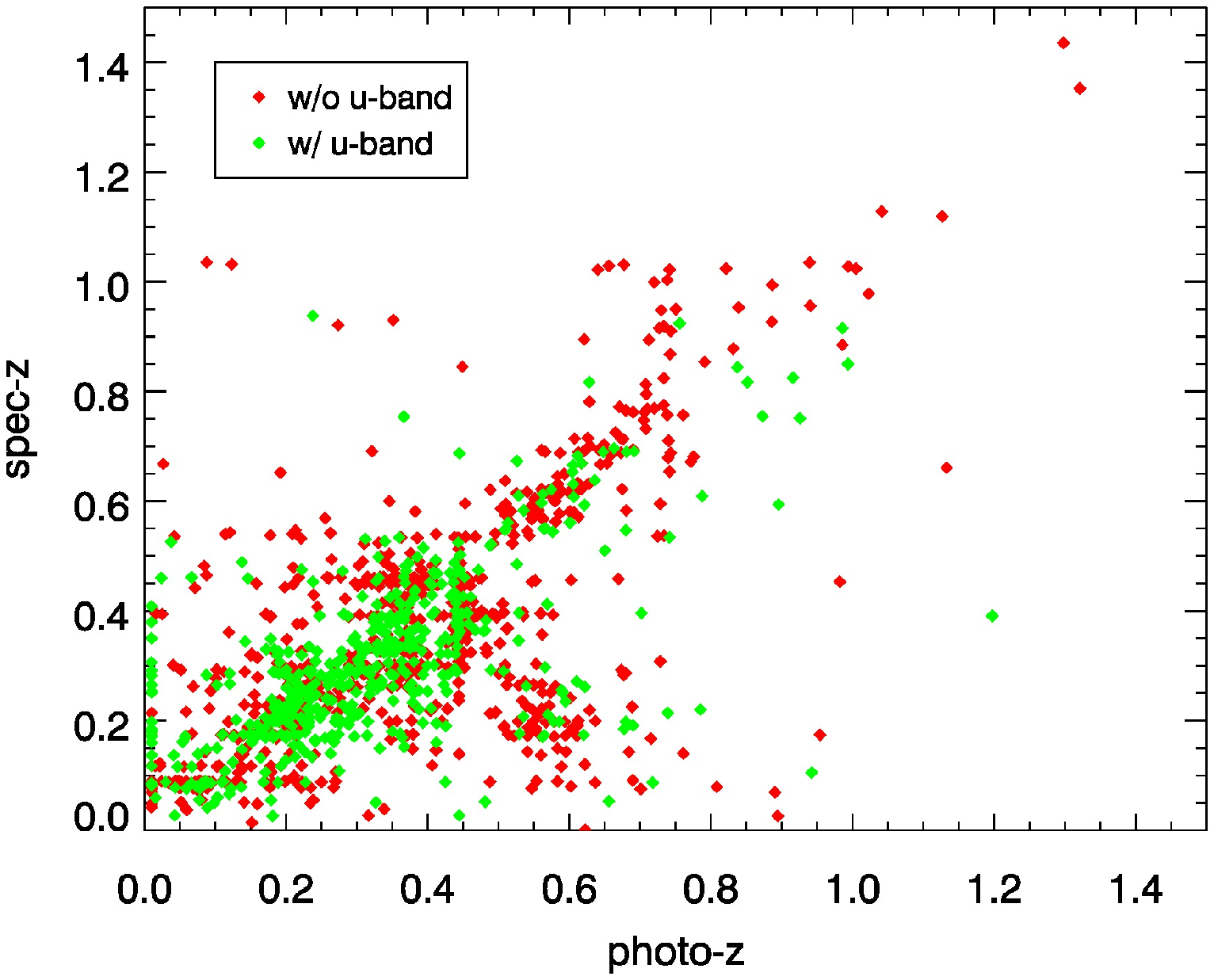}  
  \caption{
  (left) HSC three color ($g,r,i$) composite image of the NEP wide field (5.4 deg$^2$). For the first time, the entire NEP wide field was covered with deep optical data ($g\sim$27.2).\\
    (right) Photometric redshifts against spectroscopic redshifts. The sample with the CFHT $u^{*}$-band magnitude is plotted in green, while the sample without the CFHT $u^{*}$-band magnitude is plotted in red.}
  \label{fig:HSC}
  \end{center}
 \end{figure}

However, previously, we only had deep optical images in the central 0.25 deg$^2$, while previous AKARI's mid-IR data exist in much larger field of 5.4 deg$^2$. Therefore, our previous work also suffered from the cosmic variance.

To rectify the situation, we carried out an optical survey of the AKARI NEP wide field using Subaru's new Hyper Suprime-Cam \cite[HSC;][]{2012SPIE.8446E..0ZM}.
  The HSC has a field-of-view (FoV) of 1.5 deg in diameter, covered with 104 red-sensitive CCDs.
 It has the largest FoV among optical cameras on an 8m telescope, and  can cover the AKARI NEP wide field (5.4 deg$^2$) with only 4 FoV (the left panel of Fig.\ref{fig:HSC}).

Our immediate aim of the optical survey is to detect all AKARI sources in the optical, with photometry accurate enough for reliable photometric redshifts.
This allows us to determine the optical and IR luminosities (corresponding to direct and dust-obscured emission) from young stars and accreting massive black holes for a large sample representative of the cosmic history of the Universe.

Our proposal to the Subaru telescope was accepted twice (PI Goto).
In 2014, we were limited to the $r$-band observation due to unexpected unavailability of the filter stacker of HSC caused by the instrument troubles propagated from the telescope chiller trouble.
We observed in $r$ in 7 FoVs with sets of 5 point dithering pattern with the seeing of $\sim$1.5''. 

In 2016, we obtained data in the remaining $g,i,z,y$ filters in 4 FoV covering the NEP wide field. The 5 sigma limiting magnitudes are 27.18, 26.71, 26.10, 25.26, and 24.78 mag [AB] in $g, r, i, z$, and $y$-bands, respectively.
See Oi et al. in this volume for more details of the observation and data reduction.

\begin{table*}[tb]
\begin{center}
\caption{Summary of AKARI NEP survey data}\label{tab:multi}
 \footnotesize	
 \renewcommand{\arraystretch}{0.8} 
 \begin{tabular}{lccl}
\hline 
Observatory  & Band              &   Sensitivity/Number of objects/exposure time          & Area (deg$^2$)\\
\hline  
{\bfseries AKARI/IRC }   & {\bf 2.5-24$\mu$m}     &  $L15$=18.5AB  & {\bf 5.4}\\
Subaru/S-Cam  & $BVRi'z'$ & $R$=27.4AB   & 0.25\\
Subaru/FOCAS & optical spect.,   & 57 sources in NEP  &   $R\sim 24$ AB \\
MMT6m & optical spec.       & $\sim$1800 obj  &      5.4            \\
KPNO-2.1m & $J, H$  & 21.6,21.3 & 5.4\\
Maidanak 1.5m  & $B,R,I$  & $R$=23.1 & 3.4\\
KPNO2m/FLAMINGOS  & $J,H$  & $J$=21.6, $H$=21.3 & 5.4\\
WIRCAM & Y,J, K & 24AB  & 1\\
GALEX        & NUV, FUV              & NUV=26     & 1.5       \\
WSRT         & 20cm           &  $\sim$100~$\mu$Jy & 0.25\\
VLA-archive  & 10cm                  & 200~$\mu$Jy       & 5.4 \\
GMRT         & 610MHz                   & 60-80~$\mu$Jy   & 0.25        \\
Keck/Deimos         & optical spec.     & $\sim$1000 obj    & 0.25      \\
Subaru/FMOS         & near-IR spec.       & $\sim$700 obj   & 0.25       \\
   Herschel  & 100,160 $\mu$m  & 5-10 mJy & 0.5\\
        Herschel  & 250-500 $\mu$m & $\sim$10 mJy & 7.1\\
Chandra      & X-ray  &    30-80ks & 0.25\\
SCUBA2      & submm  &    1mJy  & 0.25\\
{\bfseries CFHT/MegaCam} & $u^*griz$                   &  $ r\leq25.9$AB    &  {\bfseries 4 }  \\
 {\bf Subaru/HSC}      & $g,r,i,z,y$  &    $r$=27.2(Fig.\ref{fig:HSC})  & {\bf 5.4}\\
\hline 
 \end{tabular}
\end{center}
\end{table*}

\subsection{CFHT $u^*$-band observation}
\label{sec:cfht_obs}
Subaru telescope does not have a $u^*$-band imaging capability, while it is critically important to accurately estimate photometric redshifts of low-z galaxies.
 Therefore, we obtained $u^*$-band image of the AKARI NEP wide field using the Megaprime camera of Canada France Hawaii Telescope (CFHT, PI:T.Goto). See more details of the observation and data reduction in Appendix.

\subsection{Photometric redshift}
Using CFHT $u^*$-band, and HSC $g,r,i,z,y$-bands data,
we calculated photometric redshifts with the {\ttfamily LePhare} code \citep{2007A&A...476..137A}. We used the COSMOS galaxy library for SED fitting \citep{2009ApJ...690.1236I}. Extinction law from \cite{2005ApJ...633..871C} was applied in the SED fitting. Also, we adopted the function {\ttfamily AUTO$\_$ADAPT} in {\ttfamily LePhare} to adjust magnitude zero points. In the SED fitting, we used 29 bands in maximum from the following instruments, GALEX ($FUV, NUV$), CFHT ($u^{*}$), HSC ($g, r, i, z, y$), AKARI ($N2, N3, N4, S7, S9, S11, L15, L18, L24$), WISE ([3.4],[4.6],[12],[22]), Spitzer ([3.6],[4.5],[5.8],[8.0],[24]), Herschel ($PSW, PMW, PLW$). Note not all data are available for all the AKARI sources.

 By comparing with spectroscopic redshifts, we examine the accuracy of photo-z.
 The $\sigma$ is defined to be the standard deviation of $\dfrac{\triangle z}{1+z_{s}}$ for $\dfrac{\triangle z}{1+z_{s}}$ < 0.15, where the $\triangle$ z is ${\lvert}{z_{p}-z_{s}}{\rvert}$ and  z$_{p}$ and z$_{s}$ are the photometric and spectroscopic redshifts, respectively.
 The fraction of the objects with $\dfrac{\triangle z}{1+z_{s}}>$ 0.15 is defined to be the catastrophic rate $\eta$. The main result is shown in the right panel of Fig.~\ref{fig:HSC}. The ($\sigma$, $\eta$) are (0.038, 17.6$\%$),  (0.039, 21.6$\%$), and  (0.036, 12.4$\%$) for the whole sample, the sample without $u^{*}$-band (red), and the sample with $u^{*}$-band, respectively. The $u^*$-band data improve both $\sigma$ and $\eta$.

 \subsection{Analysis}\label{sec:vmax}

We compute LFs using the 1/$V_{\max}$ method, as in \citet{GotoTakagi2010,GotoCFHT}.
 Uncertainties of the LF values stem from various factors such as fluctuations in 
 the number of sources in each luminosity bin, 
 the photometric redshift uncertainties,
 the $k$-correction uncertainties,
 and the flux errors. 
 To compute the errors of LFs we performed Monte Carlo simulations by creating 1000 simulated catalogs.
 Each simulated catalog contains the same number of sources, but we assigned a new redshift to each source, by following a Gaussian distribution centered at the photometric redshift with the measured dispersion $\Delta z/(1+z)$.
 The flux of each source is also changed; the new fluxes vary according to the measured flux error following a Gaussian distribution.

 For the 8$\mu$m and the 12$\mu$m LFs, 
 we can ignore the errors  due to the $k$-correction 
 thanks to the continuous AKARI MIR filter coverage. 
 The TIR LF errors are estimated by re-performing the SED fitting for each of the 1000 simulated catalogs.

 We did not consider the uncertainty related to cosmic variance here since our field coverage has been significantly improved.
 For our analysis, each redshift bin covers $\sim 2\times 10^7$ Mpc$^3$ of volume, which is large enough to avoid significant effect from the cosmic variance.
 See \citet{2006PASJ...58..673M} for more discussion on the cosmic variance in the NEP field.

 All the other errors described above are added to the Poisson errors for each LF bin in quadrature.

\subsection{Results: High-z IR LFs}
\label{results}

\subsubsection{The 8$\mu$m LF}\label{sec:8umlf}

Monochromatic 8$\mu$m luminosity ($L_{8\mu m}$) is known to correlate well with the TIR luminosity \citep{2006MNRAS.370.1159B,2007ApJ...664..840H,Goto2011SDSS}, especially for star-forming galaxies, because the rest-frame 8$\mu$m flux is dominated by prominent PAH (polycyclic aromatic hydrocarbon)  features such as those at 6.2, 7.7, and 8.6 $\mu$m.
 
 Since AKARI has continuous coverage in the mid-IR wavelength range, the restframe 8$\mu$m luminosity can be obtained without a large uncertainty in $k$-correction at the corresponding redshift and filter. For example, at $z$=0.375, restframe 8$\mu$m is redshifted into $S11$ filter.
 Similarly, $L15,L18W$ and $L24$ cover restframe 8$\mu$m at $z$=0.775, 1.25 and 2.
 This filter coverage is an advantage with AKARI data. Often in previous work, SED models were used to extrapolate from Spitzer 24$\mu$m fluxes, producing the largest uncertainty. 
This is not the case for the analysis present in this paper. 

 To obtain the restframe 8$\mu$m LF, we used sources down to 80\% completeness limits in each band as measured in \citet{2012A&A...548A..29K}. 
We excluded those galaxies whose SEDs are better fit with QSO templates. This removed 2\% of galaxies from the sample. 

 We used the completeness curve presented in \citet{2012A&A...548A..29K} to correct for the incompleteness of the detections. However, this correction is 25\% at maximum, since our sample is brighter than the 80\% completeness limits. Our main conclusions are not affected by this incompleteness correction.
 To compensate for the increasing uncertainty at increasing $z$, we use four redshift bins of 0.28$<z<$0.47, 0.65$<z<$0.90, 1.09$<z<$1.41,  and 1.78$<z<$2.22.  
 Within each redshift bin, we use the 1/$V_{\max}$ method to compensate for the flux limit in each filter.

\begin{figure}
 \includegraphics[height=8cm,trim={2cm 0 4cm 0},clip]{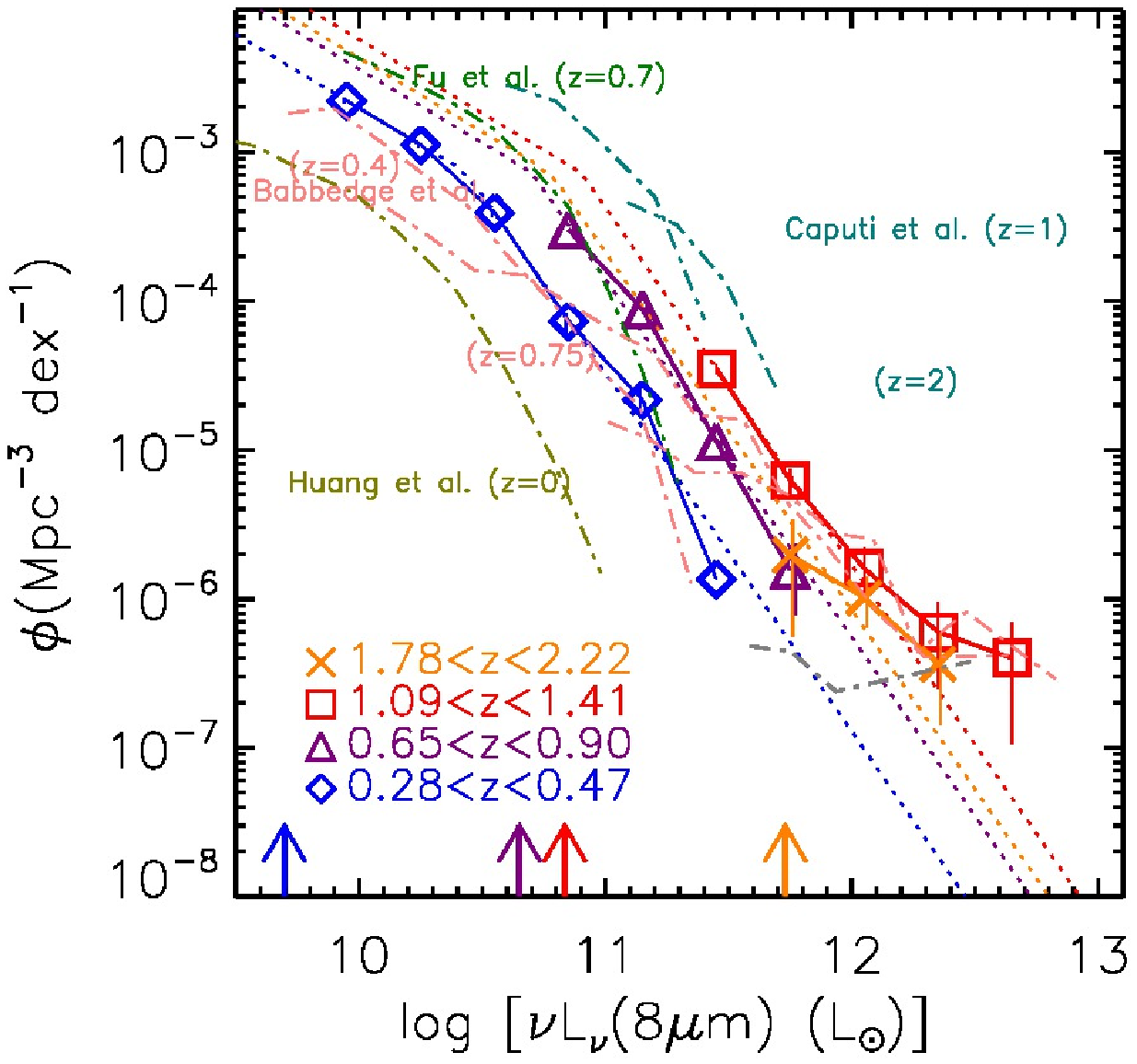}
 	\includegraphics[height=8cm,trim={2cm 0 4cm 0},clip]{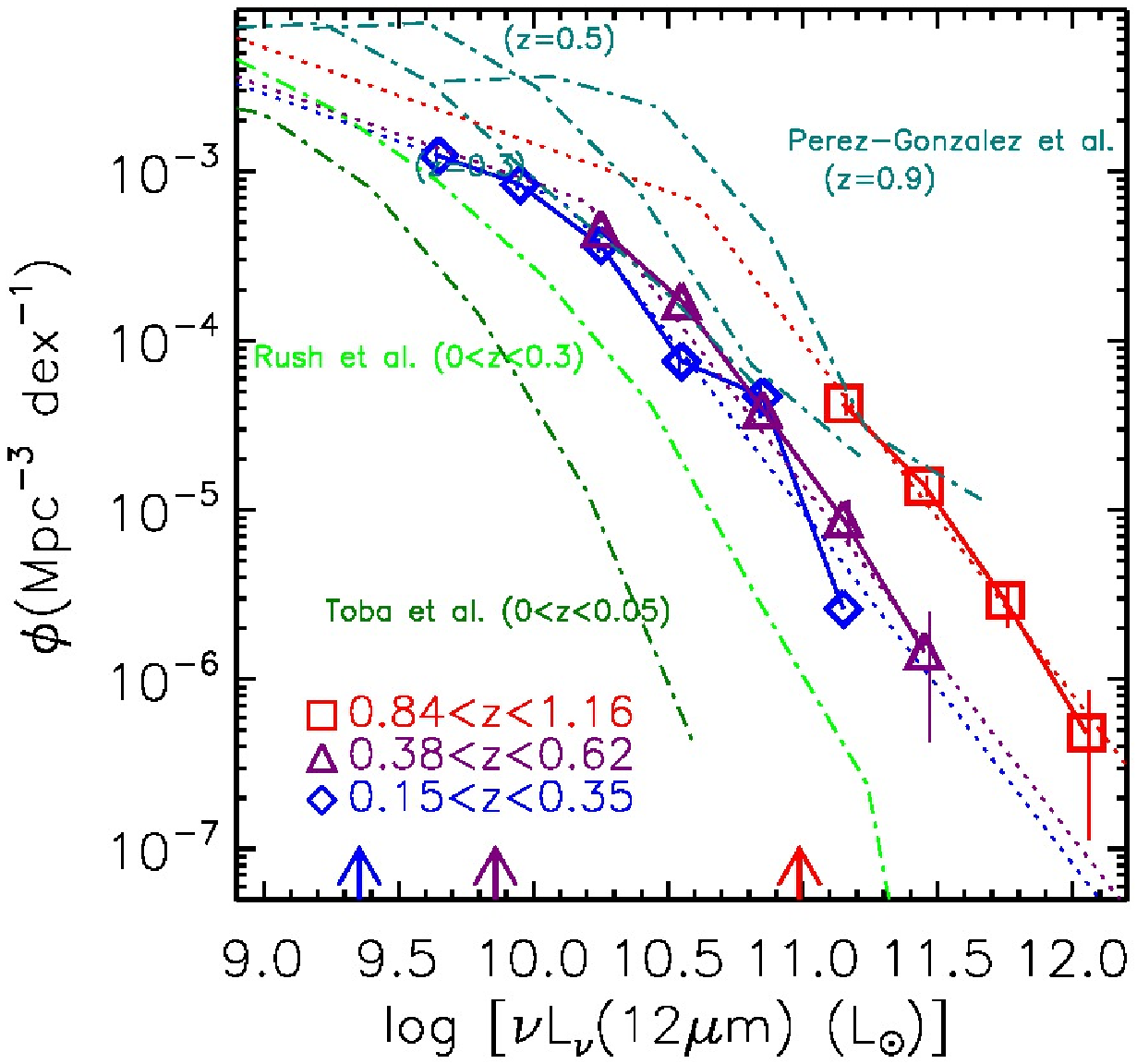}  
 \caption{
 (left)
  Restframe  8$\mu$m LFs based on the AKARI NEP wide field.
 The blue diamonds, the purple triangles, the red squares, and the orange crosses show the 8$\mu$m LFs at $0.28<z<0.47, 0.65<z<0.90, 1.09<z<1.41$, and $1.78<z<2.22$, respectively. AKARI's MIR filters can observe restframe 8$\mu$m at these redshifts in a corresponding filter. Error bars are estimated from the Monte Caro simulations ($\S$\ref{sec:vmax}).
 The dotted lines show analytical fits with a double-power law.
 Vertical arrows show the 8$\mu$m luminosity corresponding to the flux limit at the central redshift in each redshift bin.
 Overplotted are  \citet{2006MNRAS.370.1159B} in the pink dash-dotted lines, \citet{2007ApJ...660...97C} in the cyan dash-dotted lines, \citet{2007ApJ...664..840H} in the dark-yellow dash-dotted lines, and \citet{2010ApJ...722..653F} in the green dash-dotted line.\\
(right) Restframe  12$\mu$m LFs based on the AKARI NEP wide field.
Luminosity unit is logarithmic solar luminosity ($L_{\odot}$).
  The blue diamonds, the purple triangles, and the red squares show the 12$\mu$m LFs at $0.15<z<0.35, 0.38<z<0.62$, and $0.84<z<1.16$, respectively.
  Vertical arrows show the 12$\mu$m luminosity corresponding to the flux limit at the central redshift in each redshift bin.
  Overplotted are  \citet{2005ApJ...630...82P} at $z$=0.3, 0.5 and 0.9 in the dark-cyan dash-dotted lines, 
\citet{2014ApJ...788...45T} at 0$<z<$0.05 based on WISE in the dark green dash-dotted lines,
and \citet{1993ApJS...89....1R} at 0$<z<$0.3 in the light green dash-dotted lines. Note \citet{1993ApJS...89....1R} is at higher redshifts than \citet{2014ApJ...788...45T}.
 }
   \label{fig:8um}
\end{figure}

 We show the computed restframe 8$\mu$m LF in the left panel of Fig. \ref{fig:8um}. 
The arrows mark the 8$\mu$m luminosity corresponding to the flux limit at the central redshift in each redshift bin.
 Errorbars on each point are based on the Monte Carlo simulation, and are smaller than in our previous work \citep{GotoTakagi2010}.
 To compare with previous work, the dark-yellow dot-dashed line also shows the 8$\mu$m LF of star-forming galaxies at $0<z<0.3$ by \citet{2007ApJ...664..840H}, 
 using the 1/$V_{\max}$  method applied to the IRAC 8$\mu$m GTO data. 
 Compared to the local LF, our  8$\mu$m LFs show strong evolution in luminosity.

\subsection{12$\mu$m LF}\label{sec:12umlf}
 The 12$\mu$m luminosity ($L_{12\mu m}$) has been well studied through ISO and IRAS. It is known to correlate closely with the TIR luminosity \citep{1995ApJ...453..616S,2005ApJ...630...82P}. 
  As was the case for the 8$\mu$m LF, it is advantageous that AKARI's continuous filters in the mid-IR allow us to estimate restframe 12$\mu$m luminosity without much extrapolation based on SED models.
 
At targeted redshifts of $z$=0.25, 0.5, and 1, the $L15,L18W$ and $L24$ filters cover the restframe 12$\mu$m, respectively.
 The methodology is the same as for the 8$\mu$m LF; we used the sample down to the 80\% completeness limit, corrected for the incompleteness, then used the 1/$V_{\max}$  method to compute the LF in each redshift bin.
 The resulting 12$\mu$m LF is shown in the right panel of Fig. \ref{fig:8um}.
 The light green dash-dotted line shows   12$\mu$m  LF based on 893 galaxies at $0<z<0.3$ in the IRAS Faint Source Catalog \citep{1993ApJS...89....1R}. 
 The dark green dash-dotted line shows 12$\mu$m  LF at $0.006<z<0.05$ based on 223,982 galaxies from WISE sources in Table 7 of \citet{2014ApJ...788...45T}.
 Compared with these $z$=0 LFs, the 12$\mu$m LFs show steady evolution with increasing redshift.

\subsection{Total IR LFs} \label{sec:tirlf}

AKARI's continuous mid-IR coverage is also superior for SED-fitting to estimate $L_{\mathrm{TIR}}$. 
This is because for star-forming galaxies, the mid-IR part of the IR SED is dominated by the PAH emission lines,
 which reflect the SFR of galaxies \citep{1998ApJ...498..579G}, and thus, correlates well with $L_{\mathrm{TIR}}$, which is also a good indicator of the galaxy SFR. 

After photometric redshifts are estimated using the UV-optical-NIR photometry, we fix the redshift at the photo-$z$, then use the same {\ttfamily  LePhare} code to fit the infrared part of the SED to estimate TIR luminosity. 
We used \citet{2003MNRAS.338..555L}'s SED templates to fit the photometry using the AKARI bands at $>$6$\mu$m ($S7,S9W,S11,L15,L18W$ and $L24$). 

In the mid-IR, color-correction could be large when strong PAH emissions shift into the bandpass (a factor of $\sim$3). However, during the SED fitting, we integrate the flux over the bandpass weighted by the response function. Therefore, we do not use the flux at a fixed wavelength. As such, the color-correction is negligible in our process (a few percent at most).

 Galaxies in the targeted redshift range are best sampled in the 18$\mu$m band due to the wide bandpass of the $L18W$ filter \citep{2006PASJ...58..673M}. 
 Therefore, we applied the 1/$V_{\max}$  method using the detection limit at $L18W$.
 We also checked that using the $L15$ flux limit does not change our main results.
 The same \citet{2003MNRAS.338..555L}'s models are also used for $k$-corrections necessary to compute $V_{\max}$  and $V_{\min}$.
 The redshift bins used are 0.2$<z<$0.5, 0.5$<z<$0.8, 0.8$<z<$1.2,  and 1.2$<z<$1.6.  

\begin{figure}
\includegraphics[height=8cm,trim={2cm 0 4cm 0},clip]{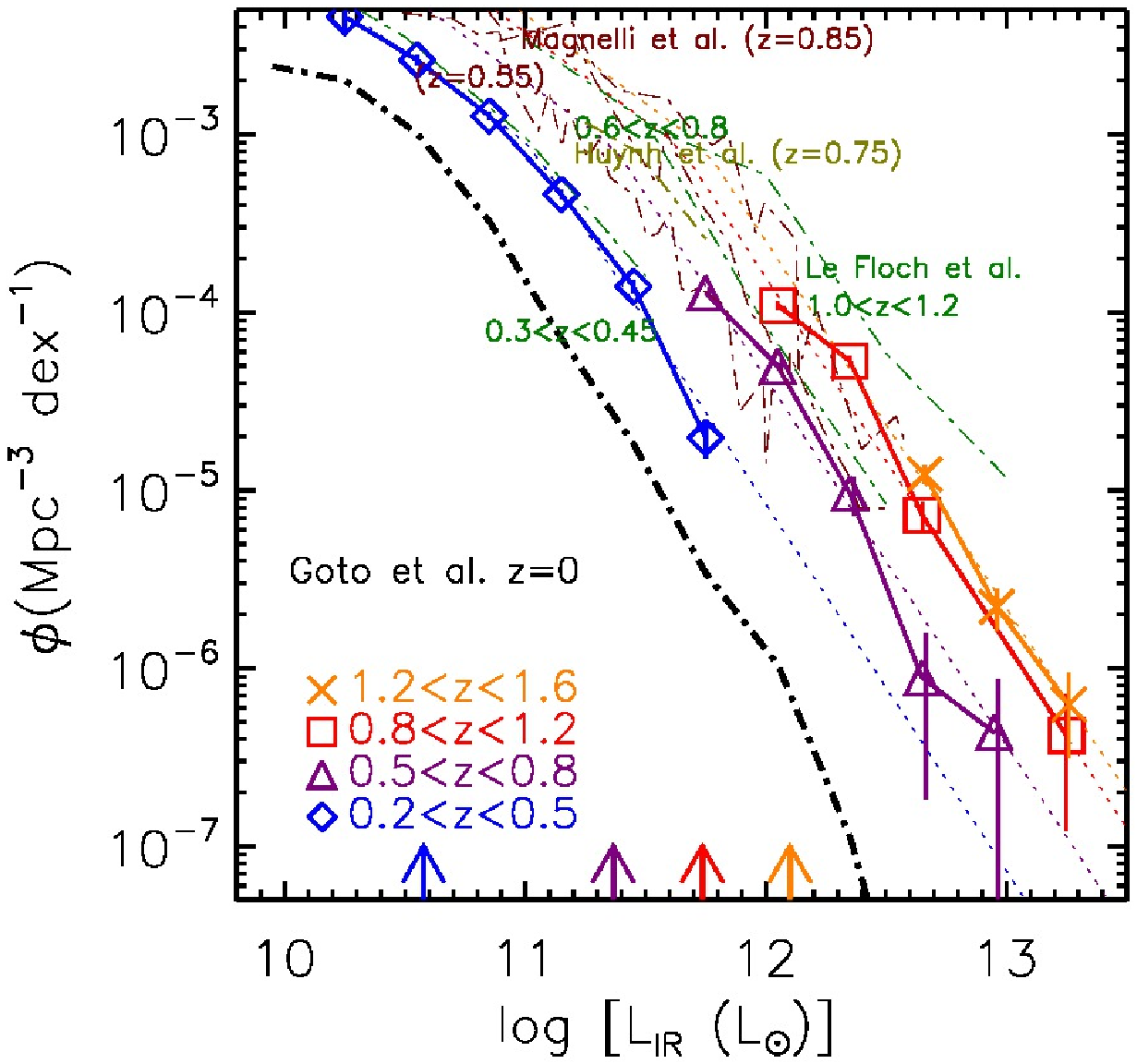}
 \includegraphics[height=8cm,trim={2cm 0 4cm 0},clip]{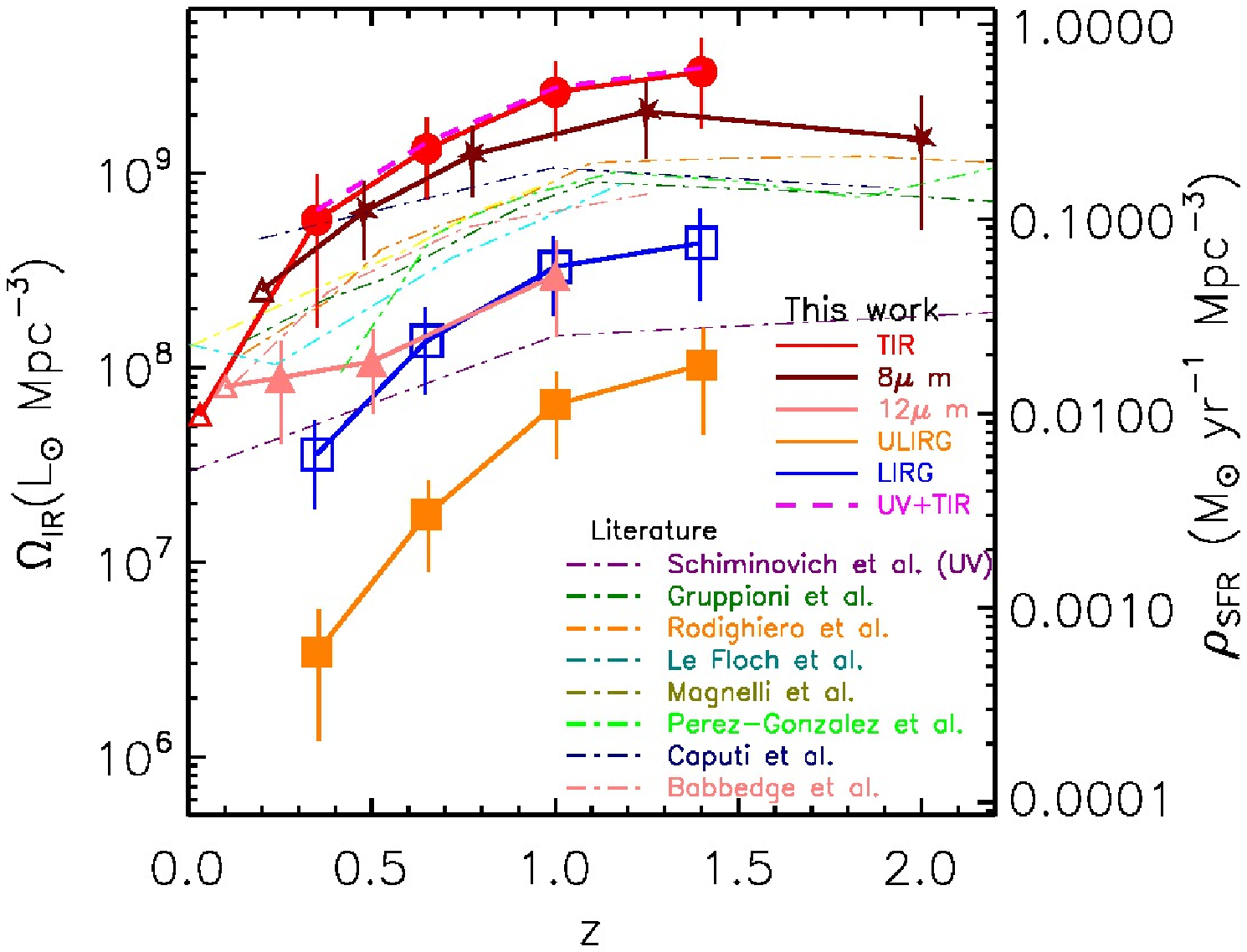} 
    \caption{(left) The TIR LFs from the SED fit.
Vertical arrows show the luminosity corresponding to the flux limit at the central redshift in each redshift bin. 
 We overplot z=0 IR LF based on the AKARI FIR all sky survey in the black dash-dot line \citep{Goto2011SDSS,Ece2017}.
 Overplotted previous studies are taken from 
\citet{2005ApJ...632..169L} in the dark-green, dash-dotted line,
\citet{2009A&A...496...57M} in the dark-red,  dash-dotted line, and 
 \citet{2007ApJ...667L...9H} in the dark-yellow,  dash-dotted line at several redshifts as marked in the figure.\\
 (right)
The evolution of the TIR luminosity density based on TIR LFs (red circles), 8$\mu$m LFs (stars), and 12$\mu$m LFs (filled triangles). 
The blue open squares and orange filled squares  are for LIRG and ULIRGs only, also based on our $L_{TIR}$ LFs. 
Overplotted dot-dashed lines are estimates from the literature: 
\citet{2005ApJ...632..169L}, 
\citet{2009A&A...496...57M}, 
\citet{2005ApJ...630...82P}, 
\citet{2007ApJ...660...97C},   
\citet{2013MNRAS.432...23G},
\citet{2010A&A...515A...8R},
and \citet{2006MNRAS.370.1159B} are in cyan, yellow, green, navy, dark green, orange, and pink, respectively.
The purple dash-dotted line shows the UV estimate by \citet{2005ApJ...619L..47S}.
The pink dashed line shows the total estimate of IR (TIR LF) and UV \citep{2005ApJ...619L..47S}. The open triangles are low-z results from \citet{Goto2011SDSS,2007ApJ...664..840H,2014ApJ...788...45T} in TIR,  8$\mu$m  and 12$\mu$m, respectively. }
\label{fig:madau}
\end{figure}

 The obtained $L_{\mathrm{TIR}}$ LFs are shown in the left panel of Fig. \ref{fig:madau}.
 The uncertainties are estimated through the Monte Carlo simulations ($\S$\ref{sec:vmax}).
 For a local benchmark, we overplot \citet{Ece2017} from the AKARI all sky survey in $\S$~\ref{142349_30Nov17}.
 The TIR LFs show a strong evolution compared to local LFs. 

\subsection{Total IR Luminosity density} \label{sec:madau}

One of the important purposes in computing IR LFs is to estimate the IR luminosity density, which in turn is an extinction-free estimator of the  cosmic star formation density \citep{1998ARA&A..36..189K}. 
 We estimate  the total infrared luminosity density by integrating the LF weighted by the luminosity.
First, we need to convert $L_{8\mu m}$ to the total infrared luminosity.

A strong correlation between $L_{8\mu m}$ and total infrared luminosity ($L_{\mathrm{TIR}}$) has been reported in the literature \citep{2007ApJ...660...97C,2008A&A...479...83B}.
Using a large sample of 605 galaxies detected in the far-infrared by the AKARI all sky survey,
\citet{Goto2011IRAS} estimated 
the best-fit relation between  $L_{8\mu m}$ and  $L_{\mathrm{TIR}}$ as 
\begin{eqnarray}\label{eq:8um}
  L_{\mathrm{TIR}} = (20\pm5) \times \nu L_{\nu,8\mu m}^{0.94\pm0.01} (\pm 44\%).
\end{eqnarray}

The  $L_{\mathrm{TIR}}$ is based on AKARI's far-IR photometry in 65,90,140, and 160 $\mu$m, and the $L_{8\mu m}$ measurement is based on AKARI's 9$\mu$m photometry.
Given the superior statistics and availability over longer wavelengths (140 and 160$\mu$m),
we used this equation to convert  $L_{8\mu m}$ into $L_{\mathrm{TIR}}$.

 The 12$\mu$m is one of the most frequently used monochromatic fluxes to estimate $L_{\mathrm{TIR}}$.
 The total infrared luminosity can be computed from the $L_{12\mu m}$ using the conversion in \citet{2001ApJ...556..562C,2005ApJ...630...82P}.

\begin{eqnarray}\label{eq:12um}
\log L_{\mathrm{TIR}}=\log (0.89^{+0.38}_{-0.27})+1.094 \log L_{12\mu m}\label{Aug  7 17:32:13 2009}
\end{eqnarray}

The 8, 12$\mu$m and total LFs are weighted by the $L_{\mathrm{TIR}}$ and integrated to obtain the TIR density.
For integration, we first fitted an analytical function to the LFs.



In this work, a double-power law is fitted to the lowest redshift LF to determine the normalization ($\Phi^{*}$) and slopes ($\alpha,\beta$). 
 For higher redshifts we do not have enough statistics to simultaneously fit 4 parameters ($\Phi^{*}$, $L^*$, $\alpha$, and $\beta$).  Therefore, we fixed the slopes and normalization at the local values and varied only $L^*$ for the higher-redshift LFs.
 Fixing the faint-end slope is a common procedure with the depth of current IR satellite surveys \citep{2006MNRAS.370.1159B,2007ApJ...660...97C}.
 The stronger evolution in luminosity than in density found by previous work \citep{2005ApJ...630...82P,2005ApJ...632..169L} also justifies this parametrization. 

 The best-fit power-laws are shown with dotted-lines in Figs. \ref{fig:8um} and \ref{fig:madau}.
 Once the best-fit parameters are found, we integrate the double power law outside the luminosity range in which we have data to estimate the TIR luminosity density, $\Omega_{\mathrm{TIR}}$.
  In the right panel of Fig. \ref{fig:madau}, we plot $\Omega_{\mathrm{IR}}$ estimated from the TIR LFs (red circles), 8$\mu$m LFs (brown stars), and 12$\mu$m LFs (pink filled triangles). All our measurements show a strong evolution as a function of redshift. 

 We also plot the contributions to $\Omega_{\mathrm{IR}}$ from LIRGs (Luminous InfraRed Galaxies; $L_{\mathrm{TIR}}>10^{11}L_{\odot}$) 
 and ULIRGs  (Ultra-Luminous InfraRed Galaxies; $L_{\mathrm{TIR}}>10^{12}L_{\odot}$, measured from TIR LFs), with the blue open squares and orange filled squares, respectively. Both LIRGs and ULIRGs show a strong redshift evolution.

\subsection*{Acknowledgments}
 TG acknowledges the support by the Ministry of Science and Technology of Taiwan through grant 105-2112-M-007-003-MY3.

\appendix
\section{Appendix: CFHT $u^{*}$-band observation and data reduction}

\begin{table}
 \centering
  \tiny	
	\caption{The configuration of extraction and photomety used for generating the CFHT $u^{*}$-band catalogue }
 \label{tab:configuration}
 \renewcommand{\arraystretch}{0.8} 
	\begin{tabular}{lccr} 

		Parameter & Value \\
		\hline
		DETECT$\_$TYPE & CCD  \\
		DETECT$\_$MINAREA & 5 \\
		DETECT$\_$MINAREA & 0 \\
		THRESH$\_$TYPE & RELATIVE \\
		DETECT$\_$THRESH & 1.5 \\
		ANALYSIS$\_$THRESH & 5 \\
		FILTER & Y \\
		FILTER$\_$NAME & gauss$\_$3.0$\_$7x7.conv \\
		FILTER$\_$THRESH & 1.0 \\
		DEBLEND$\_$NTHRESH & 32 \\
		DEBLEND$\_$MINCONT & 0.001 \\
		CLEAN & Y \\
		CLEAN$\_$PARAM & 1.0 \\
		MASK$\_$TYPE & CORRECT \\
		PHOT$\_$APERTURES & 5 \\
		PHOT$\_$AUTOPARAMS & 2.5, 3.5 \\
		PHOT$\_$PETROPARAMS & 2.0, 3.5 \\
		PHOT$\_$AUTOAPERS & 0.0,0.0  \\
		PHOT$\_$FLUXFRAC & 0.5 \\
		SATUR$\_$LEVEL & 50000 \\
		SATUR$\_$KEY & SATURATE \\
		MAG$\_$ZEROPOINT & 25.26 \\
		MAG$\_$GAMMA & 4.0  \\
		GAIN & 0.0 \\
		GAIN$\_$KEY & GAIN\\
		PIXEL$\_$SCALE & 1.0\\
		\hline
	\end{tabular}
\end{table}

 Here we summarize our CFHT  $u^*$-band observation and data reduction.
 The purpose of the $u^*$-band observation is to provide a $u^{*}$-band catalogue in AKARI NEP wide field and then estimate accurate photometric redshifts with the aid of the $u^{*}$-band magnitude. The Megaprime $u^*$-band ranges from 3235\AA{} to 4292\AA{} with the effective wavelength at 3827.2\AA{}. The observations were carried out in two time periods from 2015 May 22nd to 26th, and from 2016 July 6th to 7th (PI: T.Goto). There are 66 frames in total, which cover the 4.5 deg${^2}$ of AKARI NEP field. The exposure time is 300 second for each frame. Megaprime has 40 CCDs with 2048 $\times$ 4612 pixels for each. The pixel scale is 0.185 arcsec per pixel.  

We used the Elixir pipeline to reduce our raw $u^{*}$-band data. We masked bad pixels, bias structures and corrected for flat-fields. The pipeline had some problem with removing overscan regions, so we manually wrote a script to remove them. The zero point magnitude measured for camera runs are 25.188 and 25.121. 

To create a coadded image, we utilized AstrOmatic softwares including SExtractor, SCAMP and SWarp. First, we used SExtractor to extract sources and performed photometry. Second, SCAMP help us with astrometric calibration. In this process, we used 2MASS $J$-band observation as a reference catalogue for astrometry. Third, we coadd images from each frames by SWarp. Each pixel value is the median of every combined pixels. We include background subtraction with mesh size of 128 pixels. Last but not least, we ran SExtractor again on the final coadded image to obtain the $u^{*}$-band catalogue. The configuration of extraction and photometry is listed in Table~\ref{tab:configuration}.  We cross-matched the $u^{*}$-band catalogue with the Subaru HSC catalogue in the AKARI NEP field by matching celestial coordinate within 1 arcsec tolerance radius.

The magnitude is calculated by the following equation.
\begin{align*}
m = &-2.5 \times log(\mathrm{data\ number}) + 2.5 \times log(\mathrm{exposure\ time})+ m_{0} + K \times (\mathrm{airmass} -1)
\end{align*}
The $m_{0}$ is the zero point magnitude with value 25.121. The $K$ is a coefficient for airmass term correction, of which value is $-0.35$. Finally, we calibrated the $u^{*}$-band magnitude with the $u^{*}$-band of the AKARI NEP deep field, which is a 0.6 deg$^{2}$ sub-region of the AKARI NEP field \citep{2014A&A...566A..60O}.

\bibliography{201404_goto} 
\bibliographystyle{mnras} 
\end{document}